\documentclass[aps,pre,showpacs,floatfix,twocolumn,superscriptaddress]{revtex4-2}
\usepackage{graphicx}
\usepackage{bm,amsmath}
\usepackage{dcolumn}
\usepackage{amsfonts,amssymb}
\usepackage{chemfig}
\usepackage{diagbox}

\begin{document}
\title{Deterministic Loop Stochastic Series Expansion Algorithm for Quantum Spin Models in Magnetic Fields}

\author{Liuyun Dao}
\affiliation{Center for Advanced Quantum Studies, School of Physics and Astronomy, Beijing Normal University, Beijing 100875, China}

\author{Yan-Cheng Wang}
\affiliation{Hangzhou International Innovation Institute, Beihang University, 311115 Hangzhou, China}
\affiliation{Tianmushan Laboratory, 311115 Hangzhou, China}

\author{Hui Shao}
\email{huishao@bnu.edu.cn}
\affiliation{Center for Advanced Quantum Studies, School of Physics and Astronomy, Beijing Normal University, Beijing 100875, China}
\affiliation{Key Laboratory of Multiscale Spin Physics, Ministry of Education, Beijing 100875, China}

\date{\today}	

\begin{abstract}

The stochastic series expansion (SSE) algorithm is one of the most powerful quantum Monte Carlo methods and has been extensively applied to the study of quantum many-body systems. Its efficiency is particularly enhanced with a deterministic loop update scheme in the study of the $S=1/2$ quantum spin systems that preserve SU(2) spin rotational symmetry. Once the symmetry is broken—such as by an external field—a directed loop method is typically required, resulting in a significant reduction in efficiency. Inspired by the SSE approach developed for the quantum Ising model, we introduce a deterministic loop SSE method that is particularly suited for antiferromagnetic systems under a staggered magnetic field. This method enables separate investigations of longitudinal and transverse modes in magnetically ordered phases arising from spontaneous symmetry breaking. We benchmark the performance of our algorithm against the standard directed loop approach applied to the antiferromagnetic Heisenberg chain and demonstrate that our method substantially reduces CPU time per Monte Carlo step, thereby can outperform the directed loop algorithm in efficiency.

\end{abstract}
\maketitle

\section{Introduction}
Over the past several decades, major advances in quantum Monte Carlo (QMC) algorithms have made it possible to perform large-scale numerical studies of a wide variety of quantum many-body models. The invention of non-local loop updates has rendered QMC simulations an indispensable tool for investigating large quantum many-body systems. Compared with algorithms employing only local updates, those incorporating non-local updates offer significant advantages. Among the most prominent approaches is the method based on the power-series expansion of the partition function—referred to in the following as the stochastic series expansion (SSE)—which is both highly efficient and broadly applicable \cite{Sandvik91,Sandvik92,Syljuasen02,Syljuasen03,Alet05}. In particular, the SSE algorithm with operator-loop updates has proven to be extremely powerful in studies of quantum spin systems \cite{Wessel01,Henelius00,Henelius02,Schmidt08,Alet16} and bosonic systems \cite{Dorneich02,Schmid02,Wang16,Hesselmann16}, including cases involving external magnetic fields and chemical potentials.

In the presence of an external magnetic field, the construction rules of the SSE operator-loop updates and those of the conventional loop algorithm are, in fact, different solutions of the same set of universal equations—namely, the directed-loops equations that follow directly from the requirement of detailed balance. Therefore, the central issue lies in how to select the optimal solution of the directed-loops equations \cite{Syljuasen02,Syljuasen03}. However, when the type of magnetic field (uniform versus staggered) or its direction (longitudinal versus transverse) is changed, one must in general rederive the corresponding solutions of the directed-loops equations, a procedure that is often nontrivial.

It is normally believed that simulations of spin models in the presence of an external magnetic field must rely on directed-loops algorithms \cite{Syljuasen02,Syljuasen03,DEmidio23}, or alternatively employ purely local update schemes, which are generally less efficient\cite{Henelius00}. Motivated by quantum Monte Carlo methods for the transverse-field Ising model~\cite{Sandvik03}, we develop a deterministic-loops update algorithm based on a similar loop construction scheme, aimed at efficiently treating spin models with magnetic-field terms. As a demonstration, we apply this method to the isotropic Heisenberg chain in the presence of staggered longitudinal and transverse fields, respectively. In the weak-field regime, our algorithm significantly reduces the CPU time compared with the standard directed-loops approach. We expect that, for ferromagnetic systems under a uniform external field, our method will exhibit comparable efficiency. For antiferromagnetic systems in a uniform field, the performance of our algorithm is similar to that of the directed-loops method at weak fields; however, as the field strength increases, the directed-loops approach quickly gains a clear advantage.

Moreover, the present method avoids the difficulty of solving the directed-loops equations, features a simple implementation, and possesses broad applicability. It is therefore expected to serve as an efficient tool for simulating a wide range of quantum spin systems \cite{Wessel01,Henelius00,Henelius02,Schmidt08,Alet16} and bosonic systems \cite{Dorneich02,Schmid02,Wang16,Hesselmann16}.

\section{Staggered longitudinal magnetic field}

\begin{figure}[h]
	\centering
	\includegraphics[width=40mm]{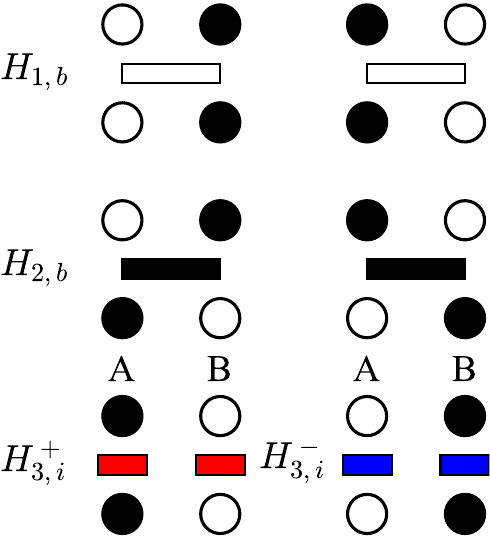}
	\caption{The six different vertices correspond to the matrix elements in Eqs.~\ref{H12b} and~\ref{H34i}. Black circles represent spin-up states, while white holes represent spin-down states. We use different colors and lengths to distinguish between the various vertices. $H_{1,b}$ and $H_{2,b}$ correspond to Heisenberg vertices, where A and B denote different sublattices. $H_{3,i}^{+(-)}$ represents magnetic vertices in which the spins are aligned with (opposite to) the magnetic field.}
	\label{schematic_bondz}
\end{figure}

As a demonstration, we focus on the introduction of a staggered magnetic field, which is crucial for breaking the spin-rotational symmetry in a finite system and is commonly used in studies of the dynamic structure factor \cite{Henelius00,Sandvik01}. Our algorithm is designed to maximize efficiency in this scenario, and it can be applied to both longitudinal and transverse magnetic fields. The Hamiltonian of the isotropic AFM Heisenberg chain under a staggered longitudinal magnetic field is defined as:
\begin{eqnarray}
H=J\sum_{\langle i,j \rangle} \boldsymbol{S}_i \boldsymbol{S}_{j}-h\sum_{i=1}^{N} (-1)^{i}S_i^z.
\end{eqnarray}
For the magnetic field term, if $i \in$ sublattice A, $sign[i]=1$, and the magnetic field causes spin $i$ to trend toward $\uparrow$; if $i \in$ sublattice B, $sign[i]=-1$, and the magnetic field causes spin $i$ to trend toward $\downarrow$. Referring to the commonly used SSE algorithm \cite{Sandvik10}, we divide the Hamiltonian into three different bond operators:
\begin{eqnarray}
	H_{1,b}&=&-S_{i(b)}^zS_{j(b)}^z+\textcolor{red}{1/4},\nonumber\\
	H_{2,b}&=&\frac{1}{2}(S_{i(b)}^{+}S_{j(b)}^{-}+S_{i(b)}^{-}S_{j(b)}^{+}),\nonumber\\
	H_{3,i}&=&(-1)^{i}S_i^{z}+\textcolor{red}{1/2+\epsilon},
\end{eqnarray}
where $H_{1,b}$ is diagonal Heisenberg operator, $H_{2,b}$ is off-diagonal Heisenberg operator, and $H_{3,i}$ is diagonal magnetic field operator. We employ vertex representations to provide an intuitive description of the bond operators and SSE configurations of the model, as illustrated in Fig.~\ref{schematic_bondz}. Constants are added to $H_{1,b}$ and $H_{3,i}$, respectively, to ensure that the series expansion is positive-definite. These constants are subtracted when calculating the energy.

The matrix elements of Heisenberg bonds are:
\begin{eqnarray}
	\langle \uparrow \downarrow \vert H_{1,b} \vert \uparrow \downarrow \rangle=1/2,\quad
	\langle \downarrow \uparrow \vert H_{1,b} \vert \downarrow \uparrow\rangle=1/2,\nonumber\\
	\langle \uparrow \downarrow \vert H_{2,b} \vert \downarrow \uparrow \rangle=1/2,\quad
	\langle \downarrow \uparrow\vert H_{2,b} \vert \uparrow \downarrow \rangle=1/2.
	\label{H12b}
\end{eqnarray}

And the matrix elements of magnetic field bonds are:
\begin{eqnarray}
	\langle \uparrow(\downarrow) \vert H_{3,i}^{+} \vert \uparrow(\downarrow) \rangle=1+\epsilon,\ 
	S_i\ aligned\ with\ h;\nonumber\\
	\langle  \downarrow(\uparrow) \vert H_{3,i}^{-} \vert \downarrow(\uparrow) \rangle=\epsilon,\ 
	S_i\ opposite\ to\ h.
    \label{H34i}
\end{eqnarray}

Then the full Hamiltonian can be written as:
\begin{eqnarray}
	H=-J\sum_{b=1}^{N_b}(H_{1,b}-H_{2,b})-h\sum_{i}H_{3,i}\nonumber \\
	+\frac{J N_b}{4}+h(1/2+\epsilon)N.
\end{eqnarray}

\begin{figure}[h]
	\centering
	\includegraphics[width=85mm]{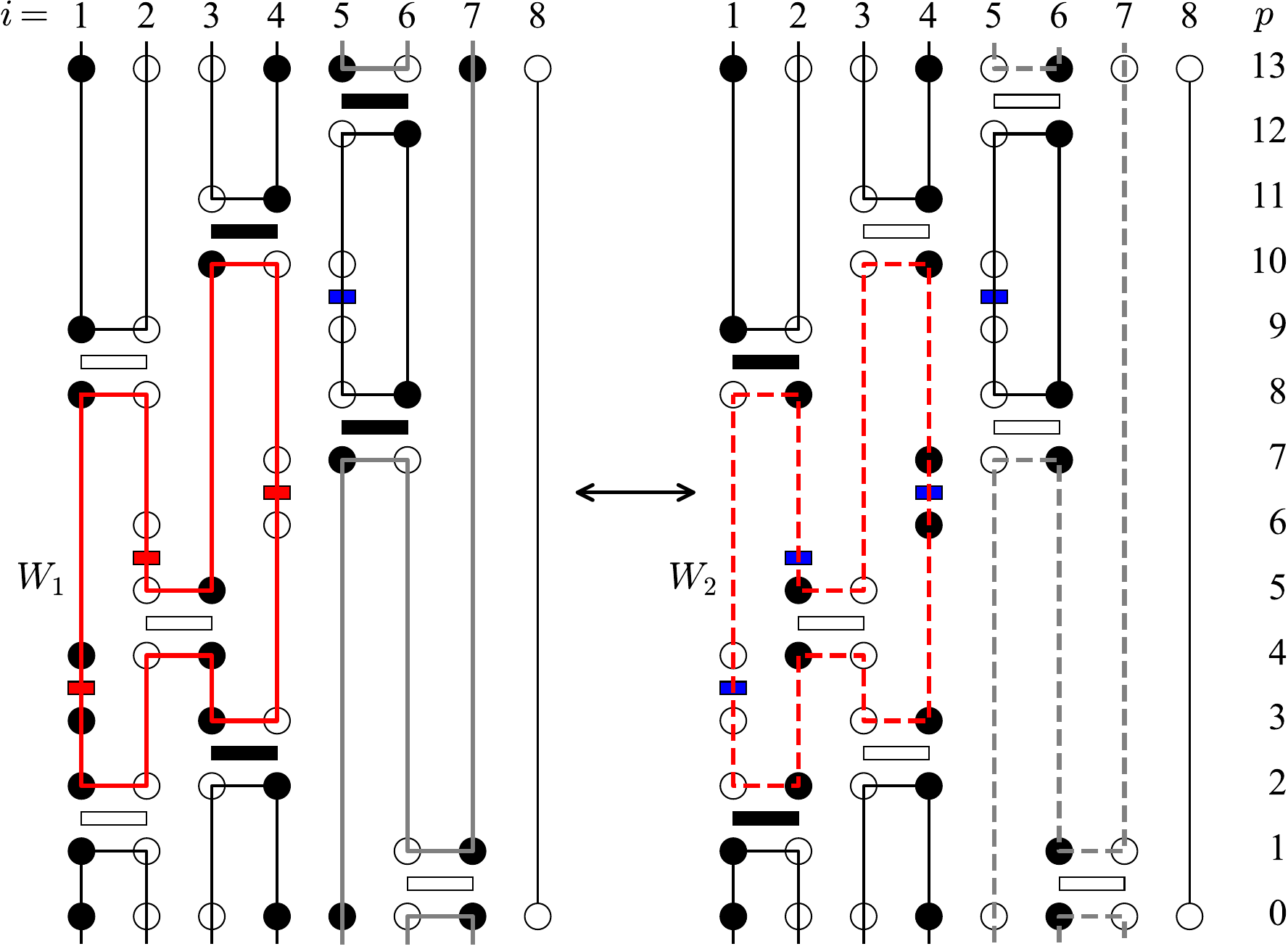}
	\caption{The schematic diagram shows an SSE configuration for an 8-spin chain. Here, $p$ denotes the index of the imaginary-time string. A closed loop is highlighted in red. The weight of the original configuration is denoted by $W_1$, while the weight after flipping the spins is denoted by  $W_2$. For loops without magnetic bonds, the probability of flipping spins is $1/2$. For loops with $n_{h}\ne 0$, the probability of flipping spins is determined by Eq.~\ref{pflip}.}
	\label{loopz}
\end{figure}

The partition function can be written as:
\begin{eqnarray}
    Z=\sum_{\alpha}\sum_{S_M}(-1)^{n_2}\frac{\beta^n(M-n)!}{M!}\langle \alpha \vert \prod_{p=0}^{M-1}H_{a(p),b(p)}\vert \alpha \rangle,
\end{eqnarray}
where $n$ is the total number of operators, $M$ is the cut-off length, and the energy is given by $E=-\langle n \rangle/\beta$. For a bipartite system, the sign factor $(-1)^{n_2}$ always positive\cite{Sandvik10}. The weight of the configuration shown in Fig.~\ref{loopz} can be written as:
\begin{eqnarray}
	W(\alpha,S_M)=(\frac{\beta J}{2})^{n_j} [\beta(1+\epsilon)h]^{n_{+}}(\beta h\epsilon)^{n_{-}}\frac{(M-n)!}{M!},\nonumber\\ 
\end{eqnarray}
here $n_j+n_{+}+n_{-}=n$. The probabilities of inserting or removing a bond are:  
\begin{eqnarray}
	P_{insert}(J)&=&\frac{\beta J N_b}{2(M-n)},\ P_{remove}(J)=\frac{2(M-n+1)}{\beta J N};\nonumber\\
	P_{insert}(h^{+})&=&\frac{(1+\epsilon)\beta h N}{(M-n)},\  P_{remove}(h^{+})=\frac{(M-n+1)}{(1+\epsilon)\beta h N};\nonumber\\
	P_{insert}(h^{-})&=&\frac{\epsilon\beta h N}{(M-n)},\ 
	P_{remove}(h^{-})=\frac{(M-n+1)}{\epsilon\beta h N}.
\end{eqnarray}

In Fig.~\ref{loopz}, we present a typical SSE configuration containing multiple loop trajectories. The construction of a loop can start from an arbitrary vertex: when the path encounters a Heisenberg vertex, a “switch-and-reverse” operation is performed, meaning that the path jumps to the other spin line connected by the bond operator while simultaneously reversing its propagation direction; when the path encounters a magnetic-field vertex, it continues straight through the operator along the original direction, while incrementing the count of visited field vertices accordingly. Eventually, the path returns to its starting point, thereby forming a closed loop.

It is worth noting that, in the directed-loops method, each time the path encounters an operator, its propagation rule must be determined by solving the directed-loopss equations. As a result, each loop-update in a QMC simulation typically (with high probability) generates a different loop structure and cluster decomposition, which is then flipped\cite{Syljuasen02,Syljuasen03}. In contrast, the deterministic-loops method treats each type of operator in a unique and fixed manner: as long as the operator string configuration is unchanged, the constructed loops remain identical in every update. In particular, when the magnetic field term is set to $h=0$, the directed-loops method reduces to the deterministic-loops method, and the cluster structure generated in each update becomes exactly the same.

Taking the configuration corresponding to the red loop in Fig.~\ref{loopz} as an example, the probability of flipping, $P(W_1 \rightarrow W_2)$ , and the probability of remaining unchanged, $P(W_1 \rightarrow W_1)$ , can be written as follows:
\begin{eqnarray}
	P(W_1 \rightarrow W_2)&=&\frac{W_2}{W_1+W_2}=\frac{1}{\left(\frac{1+\epsilon}{\epsilon}\right)^{\Delta n}+1}, \nonumber\\
	P(W_1 \rightarrow W_1)&=&\frac{W_1}{W_1+W_2}=\frac{1}{\left(\frac{1+\epsilon}{\epsilon}\right)^{-\Delta n}+1},
	\label{pflip}
\end{eqnarray}
where $\Delta n = n_l^{+} - n_l^{-}$, and $n_l^{+(-)}$ denotes the number of $H_{3,i}^{+(-)}$ operators in the closed loop. 

In the staggered-field case considered in this work, only one type of field operator can appear within a given loop, i.e., either $n_l^{+}=0$ or $n_l^{-}=0$. This is because the allowed Heisenberg bond operators always connect two sites with opposite spin orientations, thereby excluding the possibility that the two types of field operators coexist within the same loop. In contrast, for a uniform magnetic field, a single loop may contain both $H_{3,i}^{+}$ and $H_{3,i}^{-}$ operators simultaneously.

To gain a more intuitive understanding of the above flipping probabilities, let us substitute $\Delta n = 1$ and $\epsilon = 1/2$ into Eq.~\ref{pflip}. One then obtains $P(W_1 \rightarrow W_2)=1/4$ and $P(W_1 \rightarrow W_1)=3/4$. This indicates that when a loop contains operators aligned with the external magnetic field, the probability of retaining the original configuration is significantly higher than that of being flipped. This result has a clear physical interpretation: as the effect of the magnetic field increases, SSE sampling tends to favor loop configurations aligned with the field direction. For loops that do not contain any field operators, we follow the idea of the Swendsen–Wang algorithm and flip them with probability $1/2$, thereby achieving optimal update efficiency\cite{Sandvik10}.

The other components of the QMC process—including off-diagonal updates, adjustment of the cutoff, and measurement of observable quantities—are based in part on the deterministic loop algorithm developed by Sandvik in 2010 \cite{Sandvik10}, with certain modifications. One important detail is that we need to calculate the number of magnetic bonds in order to determine the flipping probability. In addition, a selection probability $P_{\text{select}}$ is introduced to decide whether to update magnetic bonds or Heisenberg bonds at a given imaginary time slice $p$ \cite{Ding17}. Although $P_{\text{select}}$ has a slight influence on the update efficiency of the QMC algorithm, it does not affect the measurement of observable quantities. The physical quantities we focus on are the order parameter
\begin{eqnarray}
	\langle m_s^z \rangle=\langle |\frac{1}{N}\sum_{i=1}^{N}(-1)^i S_i^z| \rangle.
\end{eqnarray}
Then the energy density is simply given by $e=-\langle n \rangle/(\beta N)$. 

\begin{figure}[h]
	\centering
	\includegraphics[width=65mm]{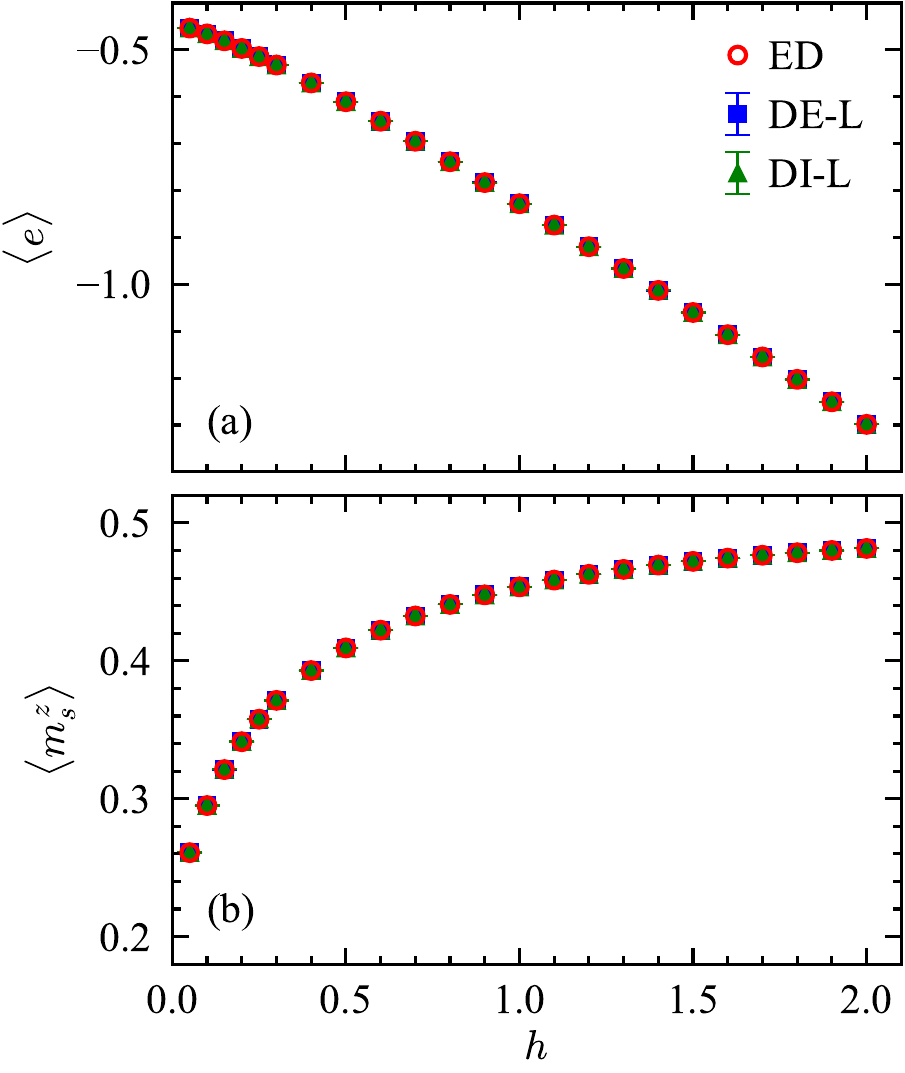}
	\caption{We compare the results of observables obtained at different magnetic field strengths for $L=12$ and $\beta=12$ using ED (red circles), deterministic-loops QMC (DE-L, blue squares), and directed-loops QMC (DI-L, green triangles): (a) energy density; (b) AFM order parameter.}
	\label{longh_ed}
\end{figure}

In Fig.~\ref{longh_ed}, we compare the results for the energy density and the order parameter obtained using Exact Diagonalization (ED), deterministic loops (DE-L) QMC, and directed loops (DI-L) QMC methods. All results show good agreement. The DI-L method is based on the widely used directed-loops algorithm developed for the XXZ model \cite{Syljuasen02,Syljuasen03}. However, the solution to the directed-loops eqnarrays differs in our case because we consider a staggered external field. By following solution B in Ref.~\cite{Syljuasen02}, we can derive a similar solution in which the bounce probabilities are zero.

To quantify the efficiency of the algorithm, we introduce the integrated autocorrelation time $\tau_{\mathrm{int}}$, which characterizes the minimum number of Monte Carlo steps (MCS) required to obtain two statistically approximately independent samples. It serves as an important measure of the efficiency of Monte Carlo (MC) sampling. For a given observable $O$, the normalized autocorrelation function is defined as
\begin{eqnarray}
A_{O}(t)=\frac{\langle O(i+t)O(i) \rangle-\langle O(i) \rangle^2}{\langle O(i)^2 \rangle-\langle O(i) \rangle^2},
\end{eqnarray}
where $i$ and $t$ denote the index of the MCS (time) and the time interval, respectively.

The autocorrelation function reflects the statistical correlation between samples at successive MCS: when $A_O(t)$ decays rapidly, neighboring samples are relatively independent and the algorithm is efficient; conversely, a slow decay indicates strong correlations and thus low efficiency. In equilibrium, the autocorrelation function typically exhibits an exponential decay, $A_O(t)\sim e^{-t/\tau_{O}}$, where $\tau_O$ is the characteristic decay time associated with the observable $O$. Based on this, the integrated autocorrelation time is defined as
\begin{eqnarray}
\tau_{\mathrm{int}}(O)=\frac{1}{2}+\sum_{t=1}^{\infty}A_{O}(t),
\end{eqnarray}
which provides a quantitative measure of the update efficiency of the Monte Carlo algorithm.

This quantity incorporates contributions from the autocorrelation function over all MCS and is a standard metric for evaluating the efficiency of MC updates. In practical calculations, one typically chooses the antiferromagnetic order parameter or the magnetic susceptibility as the observable $O$, and computes the corresponding $\tau_{\mathrm{int}}$ to assess algorithmic efficiency \cite{Syljuasen02}. A shorter integrated autocorrelation time indicates that more statistically independent configurations are generated per update, corresponding to higher efficiency. By comparing $\tau_{\mathrm{int}}$ obtained from different algorithms under the same system and parameter settings, one can directly evaluate the relative performance of various update schemes and provide guidance for optimizing SSE or directed-loops algorithms.

\begin{figure}[h]
	\centering
	\includegraphics[width=65mm]{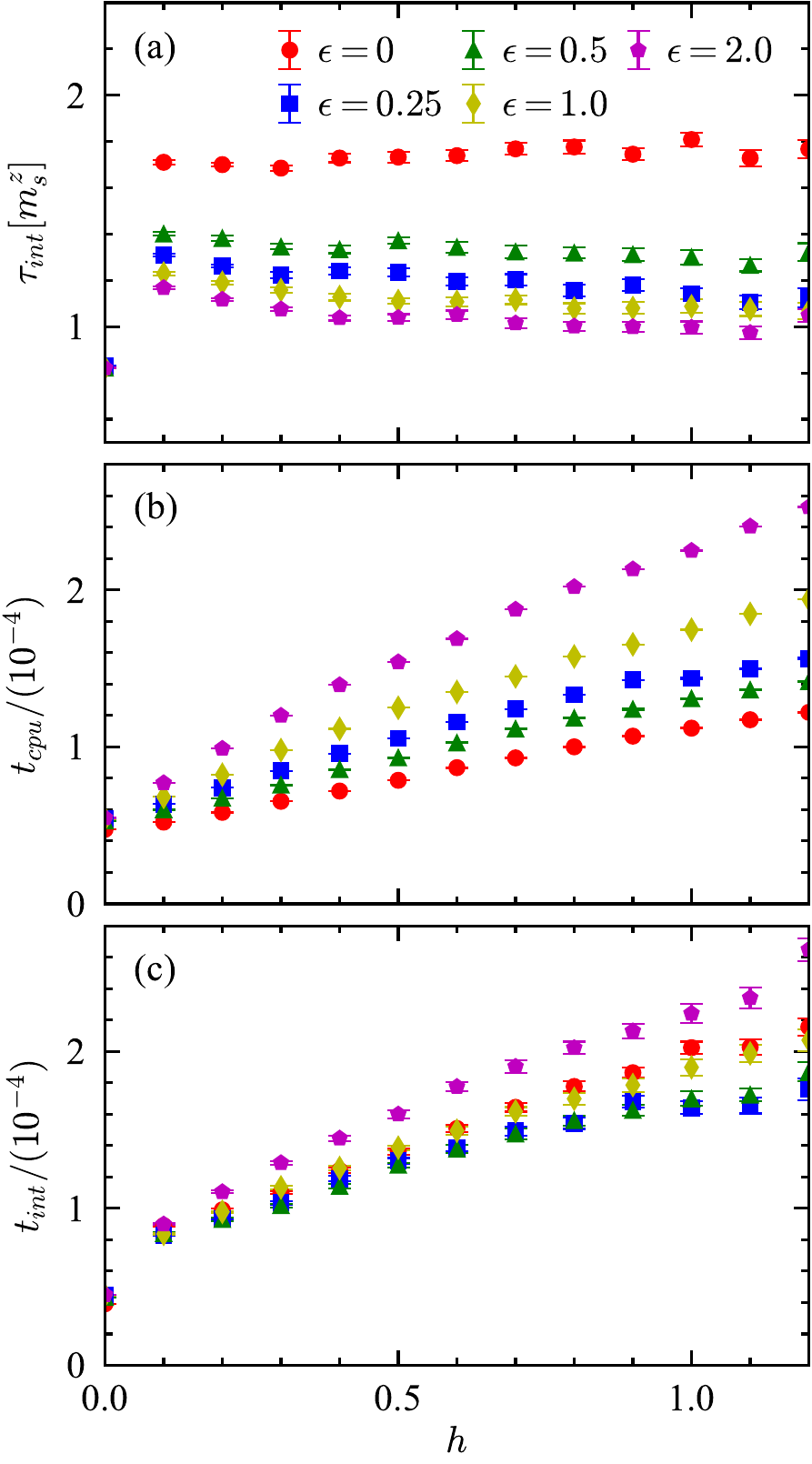}
	\caption{Results of the autocorrelation time measured using the deterministic-loops algorithm (DE-L) under different tuning parameters $\epsilon$ as a function of the external magnetic field. The system size is $L=64$, inverse temperature $\beta=16$, and the series cutoff is set to $n_{\mathrm{cut}}=500$. The symbols correspond to $\epsilon = 0.0$ (red circles), $0.25$ (blue squares), $0.5$ (green triangles), $1.0$ (yellow diamonds), and $2.0$ (purple pentagons).(a) Integrated autocorrelation time $\tau_{\mathrm{int}}[m_s^z]$ computed from the antiferromagnetic order parameter $m_s^z$;(b) CPU time per Monte Carlo step $t_{\mathrm{CPU}}$;(c) CPU time $t_{\mathrm{int}}$ required to obtain two statistically independent samples.}
	\label{time_epsilon1}
\end{figure}

In QMC simulations, we introduce an upper bound $n_{\rm cut}$ for the operator string as a cutoff for the series. As long as $n_{\rm cut} \gg \tau_{\mathrm{int}}$, this cutoff has negligible effect on the measurement of the autocorrelation time, since the autocorrelation function has essentially decayed to zero before reaching the cutoff. For the deterministic-loops algorithm (DE-L), all loops are constructed and flipped with a certain probability in each MCS, which is analogous to the Swendsen–Wang cluster update scheme \cite{Swendsen87}. In this process, the number of operators visited per Monte Carlo step, $n_{\rm vd}$, is equal to the total number of operators $n$.

In contrast, the directed-loops algorithm (DI-L) constructs and flips individual loops sequentially via off-diagonal updates, similar to the Wolff cluster algorithm \cite{Wolff89}. Therefore, during the equilibration phase, the number of loops $N_l$ needs to be adjusted to ensure that the number of operators visited per MCS approaches the maximum length $M$, i.e., $n_{\rm vd} \approx M$ \cite{Syljuasen02}. After equilibration, if an increase in the cutoff length $M$ is required, it is adjusted according to $M = \frac{4}{3} n_{\rm max}$, where $n_{\rm max}$ is the maximum number of operators encountered during the simulation. In practice, since $n_{\rm vd}(\text{DI-L}) = M > n_{\rm vd}(\text{DE-L}) = n$, the definition of a MCS is more favorable for the directed-loops algorithm, as it visits more operators per step. Nevertheless, within certain parameter regimes, the deterministic-loops algorithm can still exhibit higher sampling efficiency.

\begin{figure}[h]
	\centering
	\includegraphics[width=65mm]{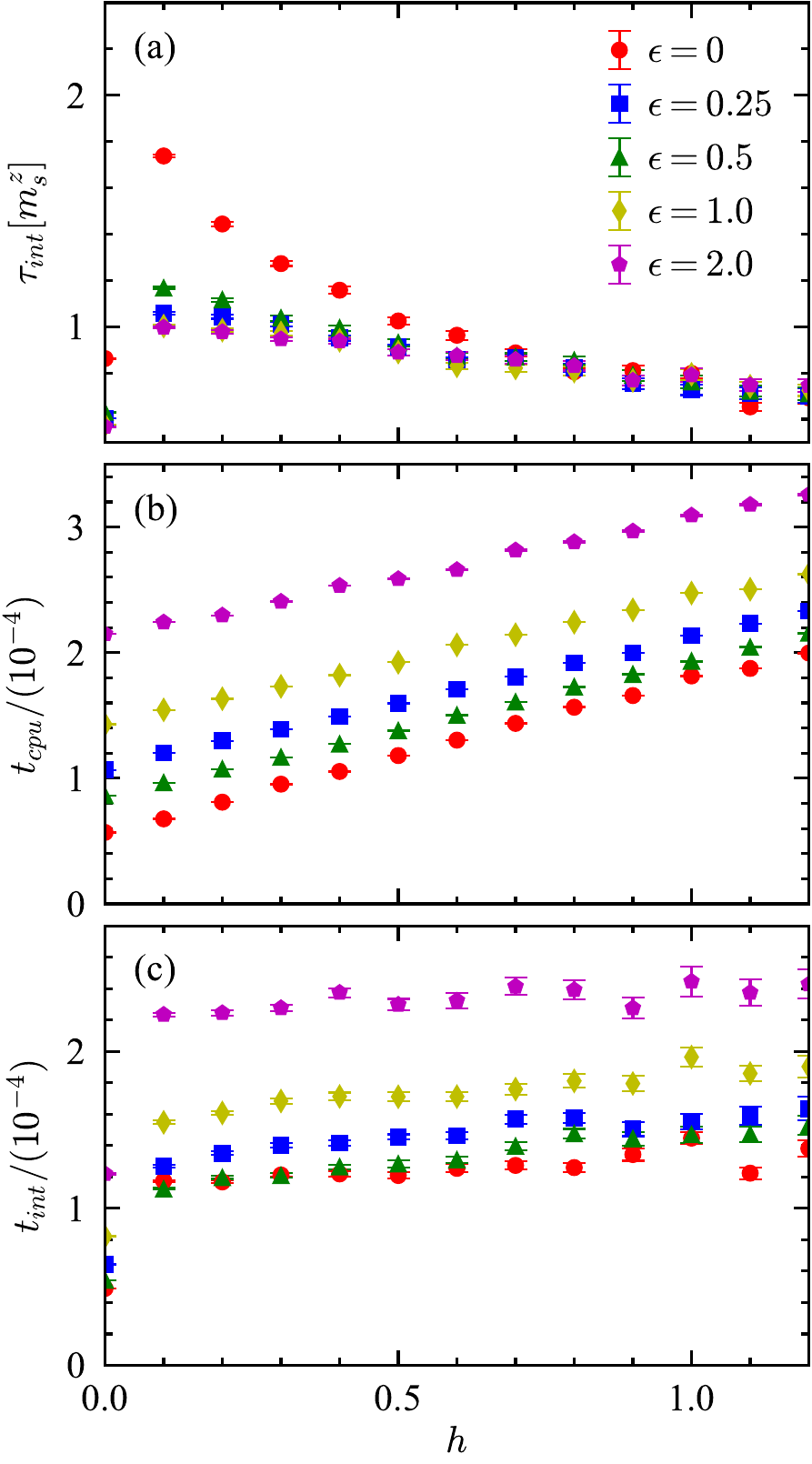}
	\caption{Results of the autocorrelation time measured using the directed-loops algorithm (DI-L) under different tuning parameters $\epsilon$ as a function of the external magnetic field. The system size is $L=64$, inverse temperature $\beta=16$, and the series cutoff is set to $n_{\mathrm{cut}}=500$. The symbols correspond to $\epsilon = 0.0$ (red circles), $0.25$ (blue squares), $0.5$ (green triangles), $1.0$ (yellow diamonds), and $2.0$ (purple pentagons).(a) Integrated autocorrelation time $\tau_{\mathrm{int}}[m_s^z]$ computed from the antiferromagnetic order parameter $m_s^z$;(b) CPU time per Monte Carlo step $t_{\mathrm{CPU}}$;(c) CPU time $t_{\mathrm{int}}$ required to obtain two statistically independent samples.}
	\label{time_epsilon2}
\end{figure}

In addition to the integrated autocorrelation time $\tau_{\mathrm{int}}$, we introduce two other quantities to evaluate the efficiency of a QMC program. The first is $t_{\mathrm{CPU}}$, the CPU time required per MCS in the simulation; the second is $t_{\mathrm{int}} = \tau_{\mathrm{int}} \times t_{\mathrm{CPU}}$, which represents the actual CPU time needed to obtain two statistically independent samples. This metric provides a comprehensive measure of algorithm performance, reflecting both statistical independence and computational cost.

In both the DE-L and DI-L algorithms, we introduce a tuning parameter $\epsilon$ to control the update efficiency of the QMC simulation. Specifically, a larger $\epsilon$ increases the complexity of the imaginary-time sequence and loop structure, which in turn raises the CPU time per MCS. At the same time, increasing $\epsilon$ can enhance the update efficiency per step, making it easier for the algorithm to generate statistically independent samples. Therefore, the choice of $\epsilon$ requires a trade-off between CPU time consumption and update efficiency. Reference \cite{Syljuasen02} provides a systematic analysis of $\epsilon$ in QMC simulations of the XXZ model under a uniform magnetic field, offering guidance for selecting an appropriate value of this parameter.

\begin{figure}[t]
	\centering
	\includegraphics[width=65mm]{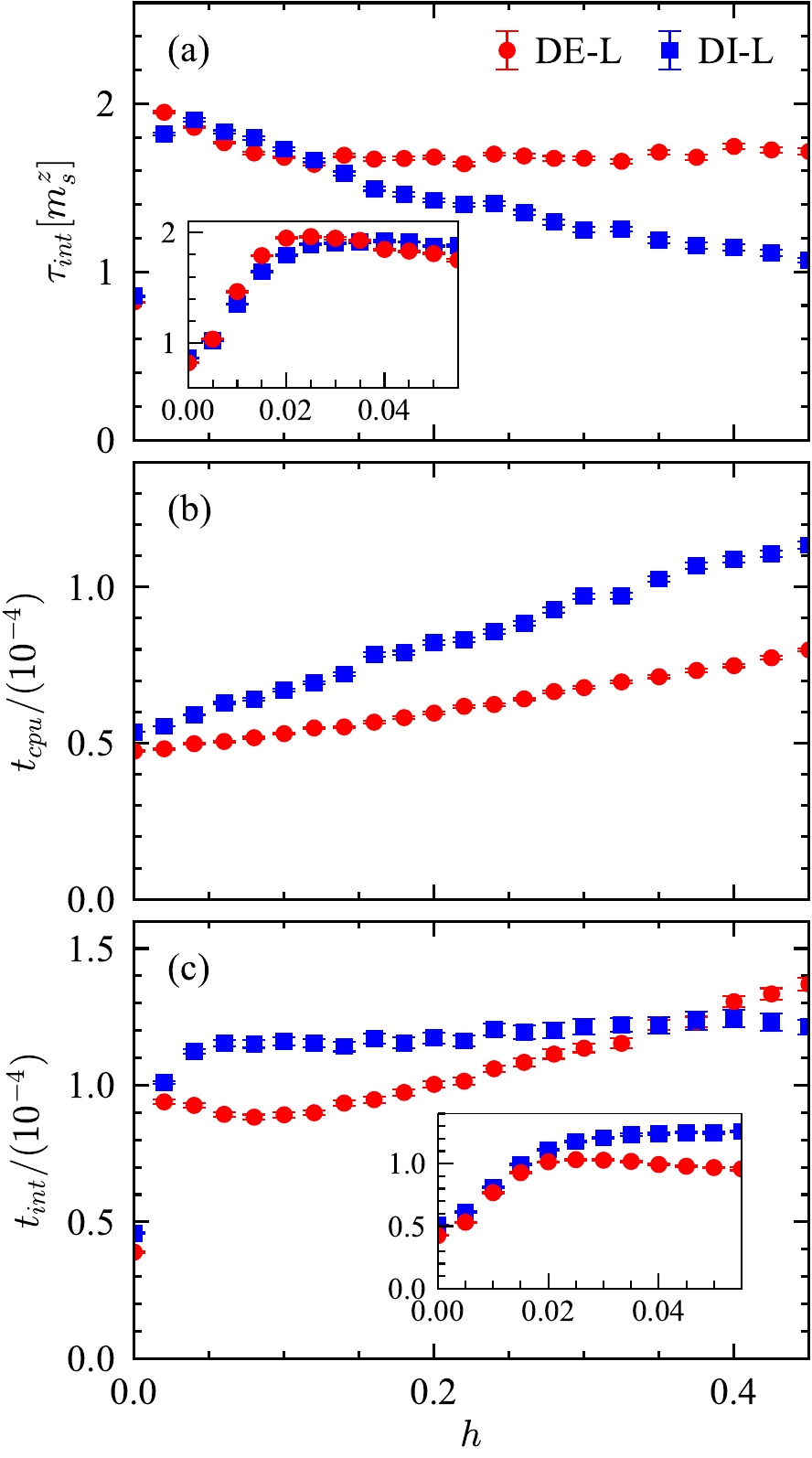}
	\caption{MC performance as a function of the external magnetic field measured using the DE-L (red circles) and DI-L ( blue squares) algorithms at $\epsilon=0$. The system size is $L=64$, inverse temperature $\beta=16$, and the series cutoff is $n_{\mathrm{cut}}=500$.(a) Integrated autocorrelation time $\tau_{\mathrm{int}}[m_s^z]$ calculated from the antiferromagnetic order parameter $m_s^z$;(b) CPU time per Monte Carlo step $t_{\mathrm{CPU}}$;(c) CPU time $t_{\mathrm{int}}$ required to obtain two statistically independent samples. Insets in panels (a) and (c) show results for a smaller magnetic field window.}
	\label{time_long_L64}
\end{figure}

\begin{figure}[t]
	\centering
	\includegraphics[width=65mm]{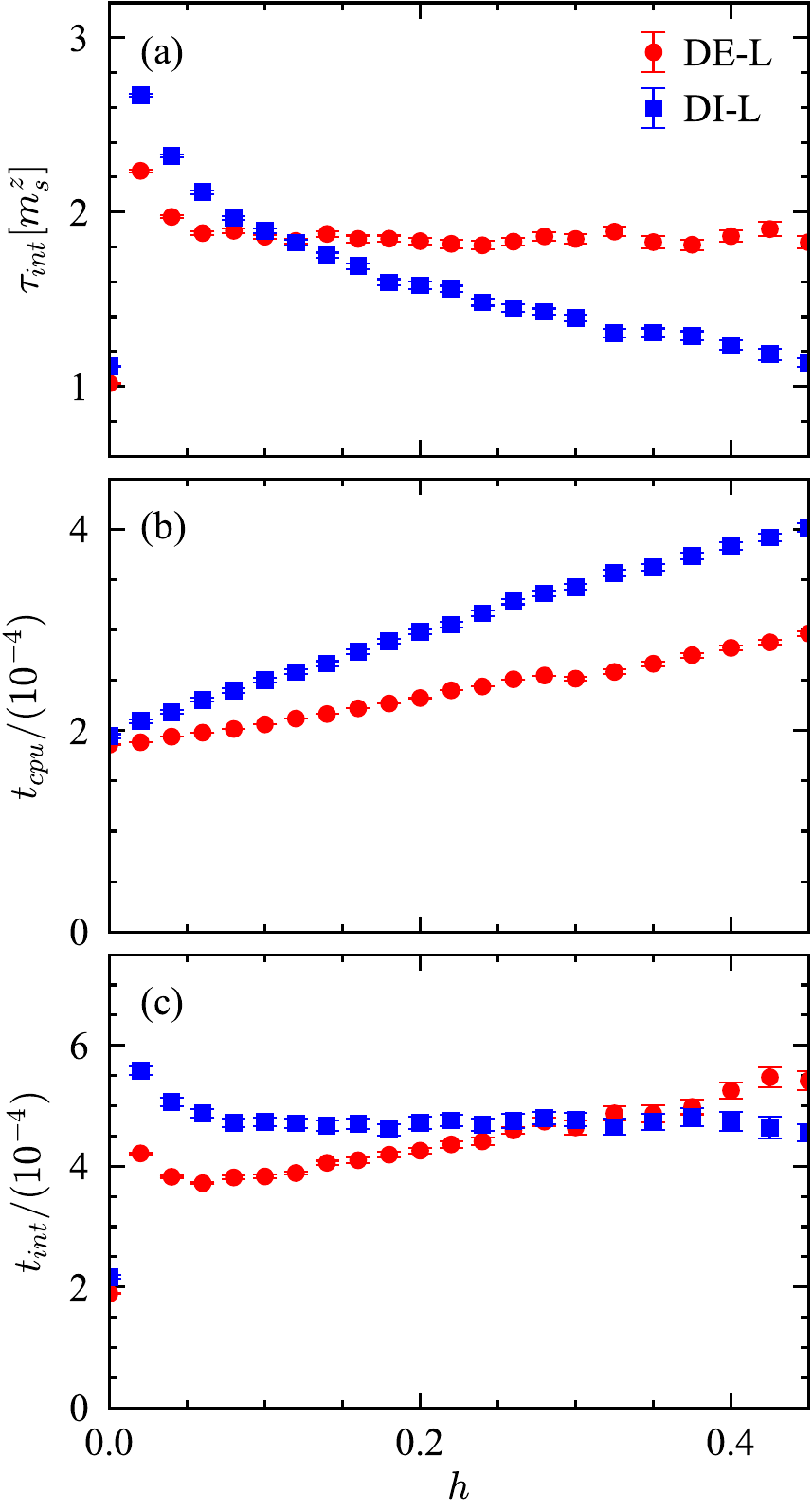}
	\caption{MC performance as a function of the external magnetic field measured using the DE-L (red circles) and DI-L ( blue squares) algorithms at $\epsilon=0$. The system size is $L=128$, inverse temperature $\beta=32$, and the series cutoff is $n_{\mathrm{cut}}=500$.(a) Integrated autocorrelation time $\tau_{\mathrm{int}}[m_s^z]$ calculated from the antiferromagnetic order parameter $m_s^z$;(b) CPU time per Monte Carlo step $t_{\mathrm{CPU}}$;(c) CPU time $t_{\mathrm{int}}$ required to obtain two statistically independent samples. }
	\label{time_long_L128}
\end{figure}

For a fair comparison, we choose the parameter $\epsilon=0$, at which both algorithms achieve their fastest simulation performance, and compare the efficiency of the DE-L and DI-L algorithms under this condition. Figures~\ref{time_long_L64} and \ref{time_long_L128} present the numerical results for system sizes $L=64$ and $L=128$, respectively.

The inset of Fig.~\ref{time_long_L64}(a) shows the behavior of the autocorrelation time in a smaller magnetic-field window. It can be seen that, for both algorithms, the autocorrelation time first increases and then decreases with increasing magnetic field. For the DE-L algorithm, the autocorrelation time reaches a maximum value of $1.95(1)$ around $h \simeq 0.02$, then decreases as the field increases, and stabilizes at approximately $1.7$ for $h \gtrsim 0.15$. In contrast, for the DI-L algorithm, the autocorrelation time forms a plateau in the range $h \in [0.02, 0.06]$, with a maximum value of about $1.9$, and then decreases continuously as the magnetic field increases further. Notably, at $h=0$, the autocorrelation times of the two algorithms are nearly identical. This result is consistent with the previous discussion: in the absence of an external magnetic field, the directed-loops algorithm reduces to the deterministic-loops algorithm in its numerical implementation, leading to nearly identical update efficiencies.

Figure~\ref{time_long_L64}(b) shows that, over the entire range of magnetic field, the CPU time consumed by the DE-L algorithm is consistently smaller than that of the DI-L algorithm. This is expected, since the DI-L algorithm is more complex in its implementation and requires visiting more loops and cluster structures in each Monte Carlo step, resulting in higher computational overhead.

From the results of $t_{\mathrm{int}}$ shown in Fig.~\ref{time_long_L64}(c), it is evident that in the magnetic-field range $h \in [0, 0.35]$, the DE-L algorithm outperforms the DI-L algorithm in overall efficiency. The CPU time required to obtain statistically independent samples differs by up to a factor of about 1.3 between the two methods. Considering that large-scale MC simulations typically require tens to hundreds of millions of MCS to achieve sufficient statistical accuracy, this efficiency advantage translates into substantial savings in CPU time in practical computations.

Figure~\ref{time_long_L128} presents the results for the system size $L=128$. Compared with $L=64$, the larger system size corresponds to a longer imaginary-time sequence, leading to significantly increased autocorrelation times and CPU costs for both algorithms. Nevertheless, in the low-field region, the DE-L algorithm still exhibits superior overall efficiency compared to the DI-L algorithm. The CPU time required to obtain statistically independent samples differs by at most a factor of about 1.3 between the two methods.

These results indicate that the efficiency advantage of the DE-L algorithm in the low-field regime remains robust even at larger system sizes. Furthermore, since the computational complexity of QMC simulations scales as $\propto \beta N$, depending only on the system size and not on the lattice dimensionality, this advantage is expected to remain robust when extended to the two-dimensional systems of interest.

It should be emphasized that the definition of a MCS adopted here is more favorable to the DI-L algorithm, since it visits more operators within each step. Meanwhile, although the chosen parameter $\epsilon=0$ lies within the optimal regime for the DI-L algorithm, it is not optimal for the DE-L algorithm (as shown in Fig.~\ref{time_epsilon1}, where the optimal range is $\epsilon \in [0.25, 0.5]$). In practical simulations, a more optimal choice of $\epsilon$ for the DE-L algorithm would further enhance its performance, leading to even greater CPU-time savings in obtaining statistically independent samples.

All benchmarks were performed on the same server within the same time period. Although slight nonphysical fluctuations in CPU time are present for a few data points, they do not affect the conclusions drawn above.

\section{Transverse magnetic field}

We now apply the magnetic field along the $x$ direction to investigate transverse quantum fluctuations in the system. The corresponding Hamiltonian can be written as
\begin{eqnarray}
    H=J\sum_{\langle i,j \rangle} \boldsymbol{S}_i \boldsymbol{S}_{j}-h\sum_{i=1}^{N}(-1)^iS_i^x,
    \label{hamhx}
\end{eqnarray}
where $S_i^x = (S^+ + S^-)/2$. For the antiferromagnetic Heisenberg model, QMC methods can only simulate a staggered transverse magnetic field; applying a uniform transverse field inevitably leads to the sign problem. Correspondingly, in the ferromagnetic model, only a uniform transverse field can be introduced \cite{Henelius00,Syljuasen03}.

In the conventional directed-loops algorithm, the single-site magnetic-field term is typically combined with the identity operator to construct an effective two-site operator with the same structure as the Heisenberg interaction. The advantage of this treatment is that, during the diagonal update, there is no need to distinguish between different operator types, and the acceptance probability depends only on the spin configuration at the ends of the operator string segment. However, the cost is that the transverse-field term introduces additional allowed spin configurations, which in turn requires the inclusion of new types of “vertices” and corresponding “processes” in the off-diagonal (loop) updates. When constructing loop diagrams, one must also consider processes in which only a single leg (incoming or outgoing) is changed while the others remain unchanged\cite{Syljuasen03}.

\begin{figure}[h]
	\centering
	\includegraphics[width=40mm]{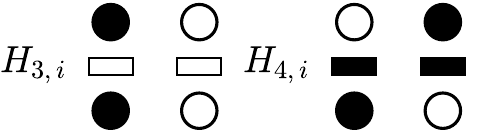}
	\caption{The schematic representation of magnetic field vertices. $H_{3,i}$ represents diagonal magnetic vertices, $H_{4,i}$ represents off-diagonal magnetic vertices.}
	\label{schematic_bondx}
\end{figure}

In contrast, the DE-L algorithm adopted here is more closely related to the treatment used for the transverse-field Ising model \cite{Sandvik03}. By explicitly separating the magnetic-field term from the interaction term, the diagonal update becomes slightly more involved, but the loop update procedure is significantly simplified. To incorporate the magnetic-field term into the generated SSE configurations, we introduce a constant operator $H_{3,i}$. The Hamiltonian is then decomposed into four mutually independent parts: the diagonal Heisenberg term $H_{1,b}$, the off-diagonal Heisenberg term $H_{2,b}$, the diagonal magnetic-field term $H_{3,i}$, and the off-diagonal magnetic-field term $H_{4,i}$, with their explicit forms given by
\begin{eqnarray}
	H_{1,b}&=&-S_{i(b)}^zS_{j(b)}^z+\textcolor{red}{1/4},\nonumber\\
	H_{2,b}&=&\frac{1}{2}(S_{i(b)}^{+}S_{j(b)}^{-}+S_{i(b)}^{-}S_{j(b)}^{+}),\nonumber\\
	H_{3,i}&=&h/2,\nonumber\\
	H_{4,i}&=&(-1)^i\frac{h}{2}(S_i^{+}+S_i^{-}).
\end{eqnarray}
We add constants to $H_{1,b}$ to ensure the series expansion is positive definite. The matrix elements of the Heisenberg terms are the same as those in Eq.~\ref{H12b}, while the matrix elements of the magnetic field terms are given by:
\begin{eqnarray}
	\langle \uparrow(\downarrow) \vert H_{3,i} \vert \uparrow(\downarrow) \rangle = h/2;\nonumber\\ 
	\langle \uparrow(\downarrow) \vert H_{4,i} \vert \downarrow(\uparrow) \rangle = h/2.
\end{eqnarray}
The schematics of $H_{3,i}$ and $H_{4,i}$ are shown in Fig.~\ref{schematic_bondx}, while those of $H_{1,b}$ and $H_{2,b}$ are the same as in Fig.~\ref{schematic_bondz}.

The Hamiltonian can be written as:
\begin{eqnarray}
	H&=-J\sum_{b=1}^{N_b}(H_{1,b}-H_{2,b})-h\sum_{i}H_{3,i}-h\sum_{i}H_{4,i}\nonumber \\
	&+\frac{J N_b}{4}+\frac{Nh}{2}.
\end{eqnarray}

The partition function can be written as:
\begin{eqnarray}
	Z=\sum_{\alpha}\sum_{n=0}^{\infty}\sum_{S_n}(-1)^{n_2+\frac{n_4}{2}}\frac{\beta^n}{n!}\langle \alpha \vert \prod_{p=1}^{n}H_{a(p),b(p)}\vert \alpha \rangle,\nonumber\\ 
\end{eqnarray}
where $n=n_1+n_2+n_3+n_4$. The probabilities of inserting or removing a magnetic bond are:  
\begin{eqnarray}
	P_{insert}(h)=\frac{\beta h N}{2(M-n)},\nonumber \\ P_{remove}(h)=\frac{2(M-n+1)}{\beta h N}. 
	\label{ptrans}
\end{eqnarray}
It should be emphasized that, in the diagonal update, only the diagonal Heisenberg operator $H_{1,b}$ and the constant operator $H_{3,i}$ are allowed to be inserted. The off-diagonal Heisenberg operator $H_{2,b}$ and the transverse-field operator $H_{4,i}$ are instead generated naturally during the subsequent loop-update process.

\begin{figure}[h]
	\centering
	\includegraphics[width=85mm]{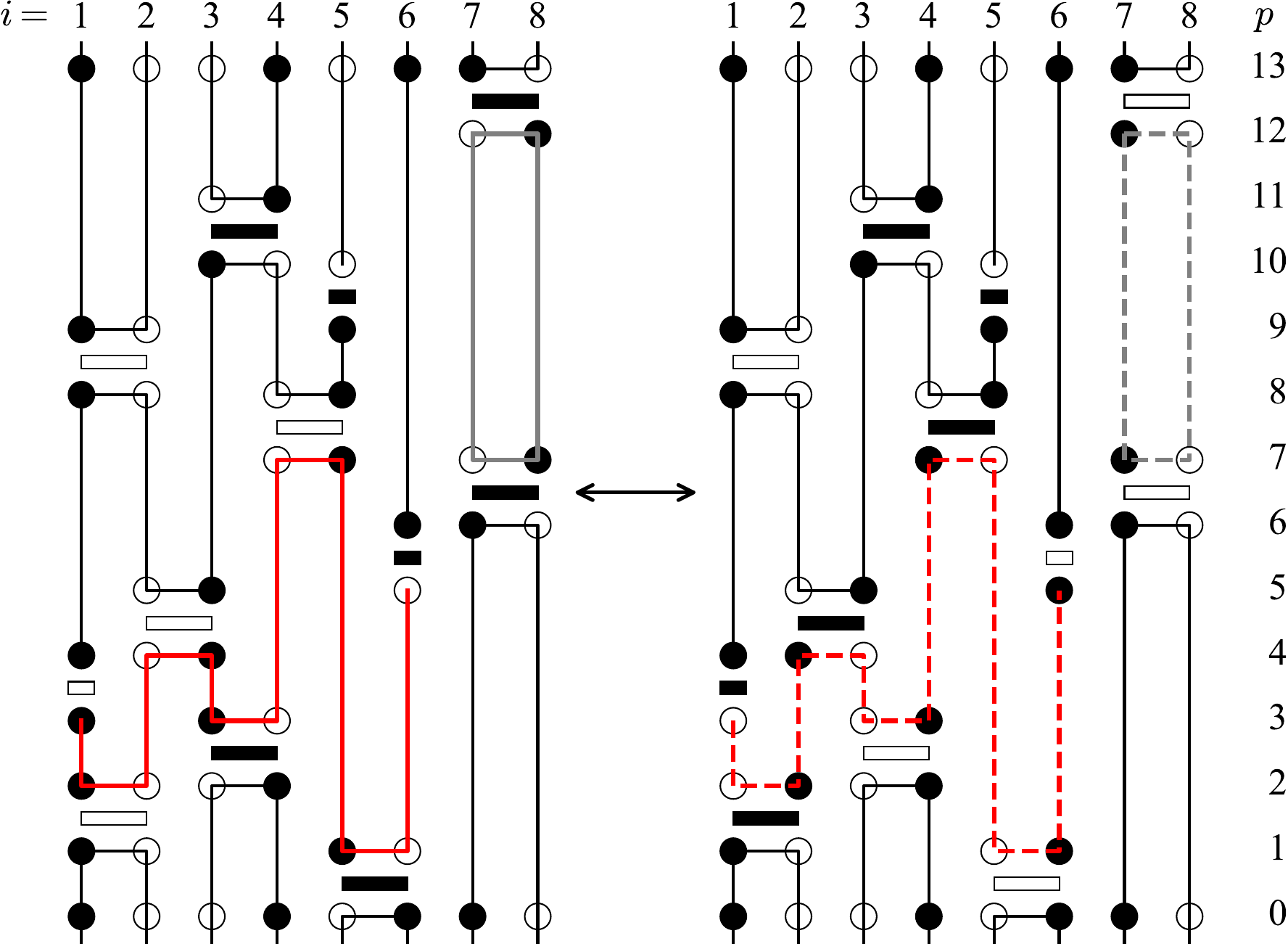}
	\caption{The schematic diagram shows an SSE configuration for an 8-spin chain. Here, $p$ denotes the index of the imaginary-time string. The magnetic field term cuts off the loops into independent clusters, whereas a closed loop can form in the absence of the magnetic field term. The weights before and after flipping the clusters are the same; therefore, the flip probability $P_{flip}=1/2$. The spins of sublattice B contribute a minus sign to ensure the positivity of the partition weight.}
	\label{clusterx}
\end{figure}

In Fig.~\ref{clusterx}, we present an example of an SSE configuration containing multiple loop trajectories. When a path encounters a magnetic-field vertex during its propagation, it terminates immediately; therefore, clusters containing magnetic-field vertices cannot form closed loops. Since the configurations before and after the loop update have identical weights, the situation is entirely analogous to that of the pure Heisenberg model: flipping each cluster with probability $1/2$ yields optimal update efficiency \cite{Sandvik10}. As illustrated by the loop highlighted in red in Fig.~\ref{clusterx}, if two clusters separated by the same magnetic-field operator are treated differently during a single loop update—one being flipped while the other remains unchanged—the conversion between the constant operator $H_{3,i}$ and the off-diagonal transverse-field operator $H_{4,i}$ is naturally realized.

It is worth noting that, although the cluster shown in the left panel of Fig.~\ref{clusterx} contains an odd number of off-diagonal operators (seven black rectangles in the figure), the overall sign factor in the partition function, $(-1)^{n_2+\frac{n_4}{2}} = (-1)^6$, remains positive. This is because, for a staggered transverse field, only the magnetic-field terms located on the $B$ sublattice contribute to the sign (site 6 in the figure), while those acting on the $A$ sublattice always carry positive weight.

For the measurement of most physical observables, the implementation of the model with a transverse field is essentially the same as that of the standard Heisenberg model or the model with a longitudinal field. Moreover, since the staggered transverse magnetization itself appears explicitly in the Hamiltonian, its measurement can be carried out analogously to that of the energy, and can be rewritten as \cite{Henelius00}:
\begin{eqnarray}
	\langle m^x_s \rangle&=&\frac{1}{N}\langle \sum_{i=1}^{N} (-1)^i S_i^x \rangle=\frac{1}{N} \langle H_{4,i} \rangle \nonumber \\ 
	&=&\sum_{\alpha}\sum_{n=0}^{\infty}\sum_{S_n}\frac{\beta^n}{n!}\langle \alpha \vert H_{4,i} \prod_{p=1}^{n}H_{a(p),b(p)}\vert \alpha \rangle \nonumber \\
	&=&\frac{\langle n_4 \rangle}{N\beta h}.
\end{eqnarray}

\begin{figure}[h]
	\includegraphics[width=65mm]{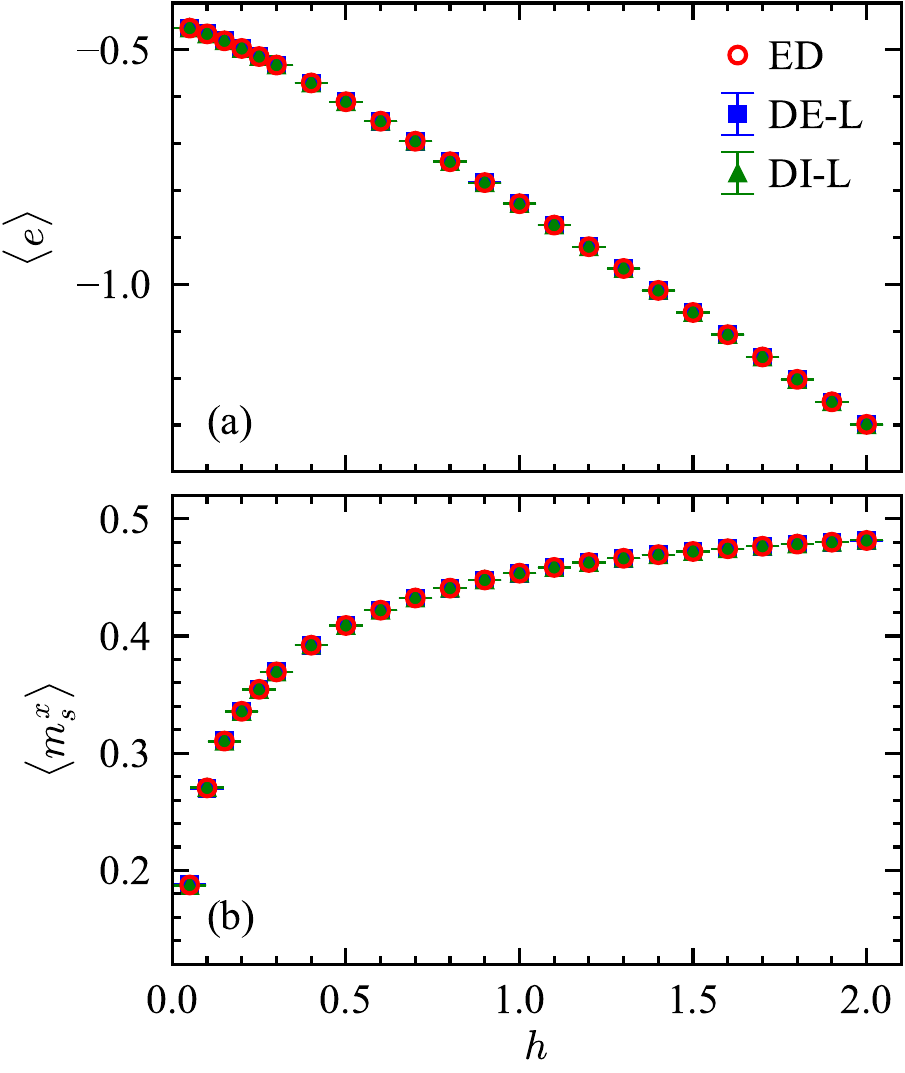}
	\caption{The results of observables obtained at different magnetic field strengths for $L=12$ and $\beta=12$ using ED (red circles), DE-L QMC (blue squares), and DI-L QMC (green triangles): (a) energy density $e$; (b) antiferromagnetic order parameter along the $x$ direction $m_s^x$.}
	\label{transh_ed}
\end{figure}

In Fig.~\ref{transh_ed}, we compare the energy density and the magnetization along the $x$ direction $m_s^x$ obtained using ED, DE-L QMC, and DI-L QMC. It can be seen that all three methods yield highly consistent results within the parameter range considered.

It should be noted that the directed-loops method employed here is based on the algorithm developed for the transverse-field XXZ model\cite{Syljuasen03}. In the parameter regime studied in this work, the bounce probability is zero, indicating that the directed-loops algorithm operates at optimal update efficiency.

\begin{figure}[h]
	\centering
	\includegraphics[width=65mm]{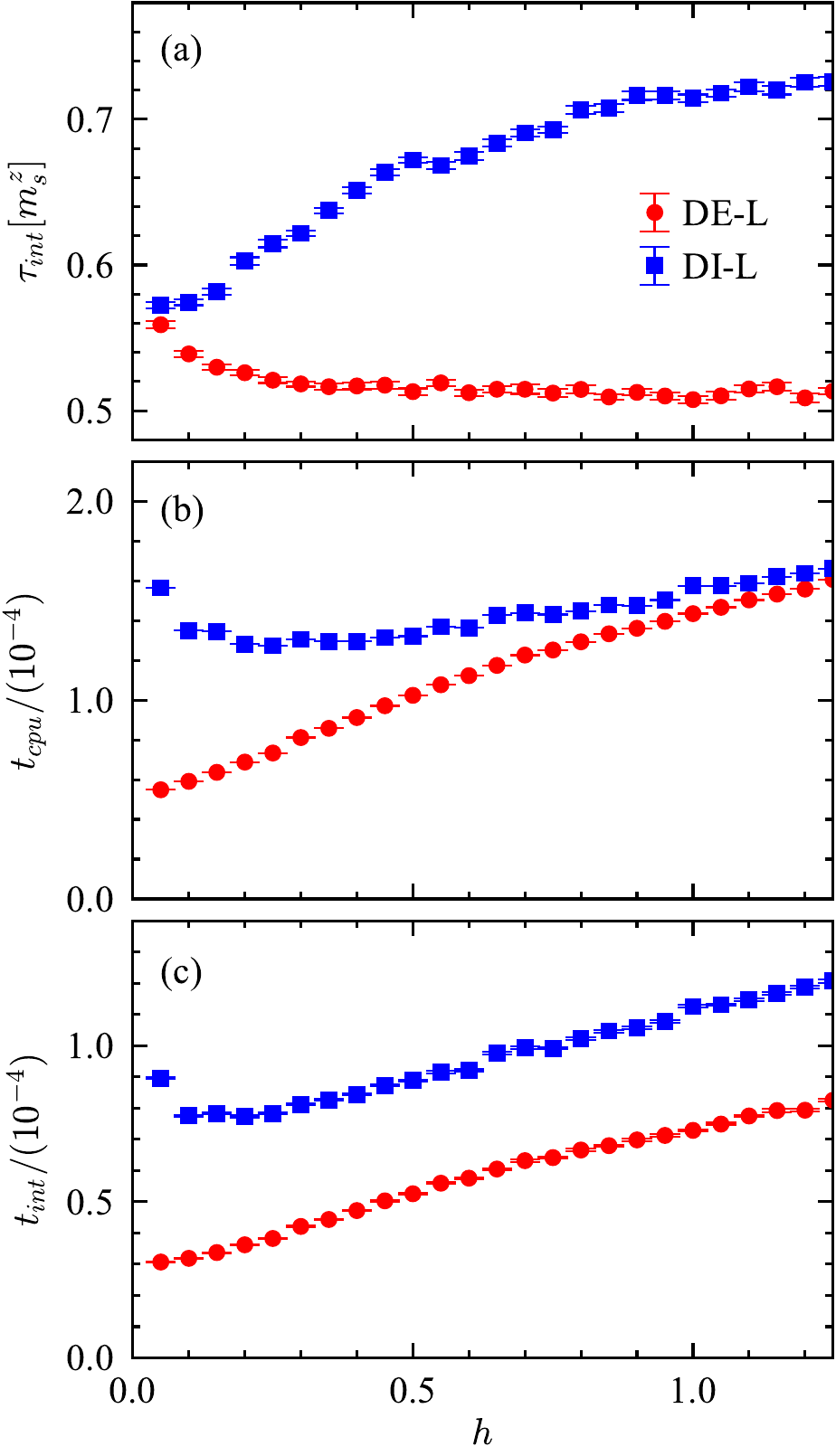}
	\caption{MC performance as a function of the external magnetic field measured using the DE-L (red circles) and DI-L (blue squares) algorithms. The system size is $L=64$, inverse temperature $\beta=16$, and the series cutoff is $n_{\mathrm{cut}}=500$. (a) Integrated autocorrelation time $\tau_{\mathrm{int}}[m_s^z]$ calculated from the antiferromagnetic order parameter $m_s^z$;(b) CPU time per Monte Carlo step $t_{\mathrm{CPU}}$;(c) CPU time $t_{\mathrm{int}}$ required to obtain two statistically independent samples.}
	\label{time_trans_L64}
\end{figure}

\begin{figure}[h]
	\centering
	\includegraphics[width=65mm]{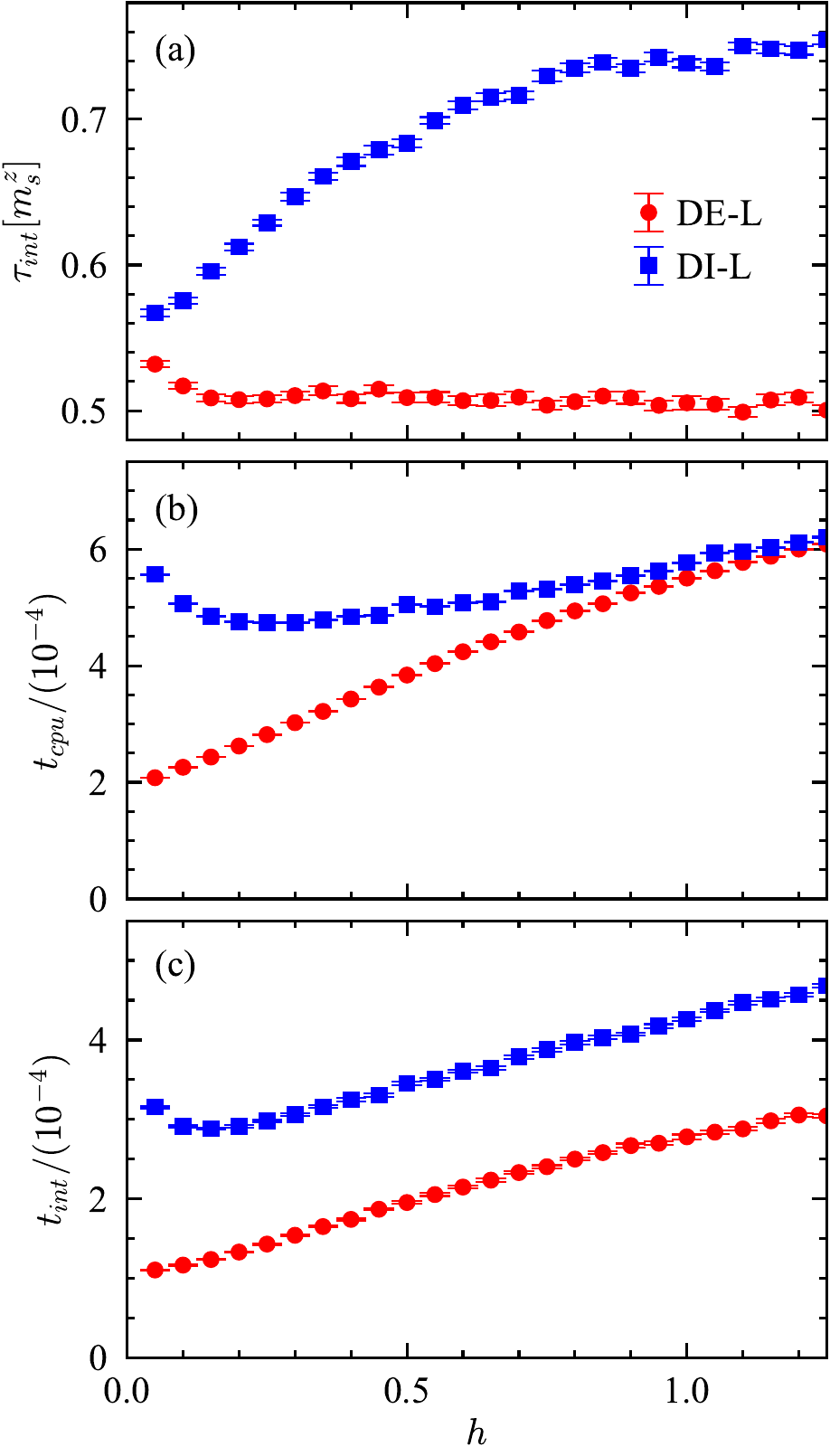}
	\caption{MC performance as a function of the external magnetic field measured using the DE-L (red circles) and DI-L (blue squares) algorithms. The system size is $L=128$, inverse temperature $\beta=32$, and the series cutoff is $n_{\mathrm{cut}}=500$. (a) Integrated autocorrelation time $\tau_{\mathrm{int}}[m_s^z]$ calculated from the antiferromagnetic order parameter $m_s^z$;(b) CPU time per Monte Carlo step $t_{\mathrm{CPU}}$;(c) CPU time $t_{\mathrm{int}}$ required to obtain two statistically independent samples.}
	\label{time_trans_L128}
\end{figure}

Since the physical quantity of interest is the transverse imaginary-time correlation function $\langle S_i^z(\tau) S_j^z(0) \rangle$ in the presence of a transverse field, we continue to use the antiferromagnetic order parameter along the $z$ direction, $m_s^z$, as the observable to compare the autocorrelation times of the two algorithms. Figures~\ref{time_trans_L64} and \ref{time_trans_L128} present the results for system sizes $L=64$ and $L=128$, respectively. It can be seen that the data for different system sizes are in excellent agreement. For the DE-L algorithm, the autocorrelation time decreases gradually with increasing magnetic-field strength, whereas for the DI-L algorithm, it increases as the magnetic field is enhanced.

The CPU time required for a single QMC update in both algorithms increases with increasing magnetic field, although the growth is more pronounced for the DE-L algorithm. When the magnetic field exceeds $h \gtrsim 1.2$, the CPU time per update for the DE-L algorithm surpasses that of the DI-L algorithm. Nevertheless, within the field range considered ($h \in [0.0, 1.2]$), the DI-L algorithm consistently requires more CPU time than the DE-L algorithm to obtain two statistically independent samples, by approximately a factor of 1.5. Given that $t_{\mathrm{int}}$ exhibits an almost linear increase with the magnetic field, this conclusion is unlikely to change even at larger field strengths.

In summary, under a transverse field, the overall efficiency of the DE-L algorithm is superior to that of the DI-L algorithm. It should be noted, however, that due to the higher degree of freedom in the directed-loops equations for the transverse-field case, the particular solution adopted in this work is not guaranteed to yield the optimal update efficiency, as also discussed in Ref.~\cite{Syljuasen03}. In principle, the efficiency of the DI-L algorithm could be significantly improved by constructing more optimal loop-update schemes and identifying more suitable solutions to the directed-loops eqnarrays. However, this is not an urgent task in practice, since the DE-L algorithm already demonstrates sufficiently high efficiency and is also simpler to implement.

\section{Conclusion}

Based on the SSE quantum Monte Carlo method, we developed a deterministic-loops algorithm applicable to the Heisenberg model in the presence of an external magnetic field, and applied it to the isotropic Heisenberg model with staggered longitudinal and transverse fields as benchmark tests. In the case of a staggered longitudinal field, the deterministic-loops algorithm significantly reduces computational time compared to the standard directed-loops method in the weak-field regime. For the staggered transverse-field case, it demonstrates higher update efficiency over a wide range of magnetic fields. In addition, the deterministic-loops approach avoids complex update operations and the need to solve directed-loops equations, making it simpler both in theoretical formulation and in practical implementation. Therefore, this method offers clear advantages in terms of learning cost and its extension to a broader class of quantum many-body systems, such as quantum spin systems and bosonic models.

\end{document}